\documentclass[superscriptaddress,preprint,12pt]{revtex4-1}

\usepackage{amsmath}    
\usepackage{graphicx}   
\usepackage{verbatim}   
\usepackage{color}      
\usepackage{subfigure}  
\usepackage{hyperref}   
\usepackage{amssymb}
\usepackage{amsthm}
\usepackage{amsfonts}
\usepackage{mathrsfs} 
\usepackage{setspace} 
\usepackage{caption} 
\usepackage[version=4]{mhchem}
\usepackage{siunitx}
\usepackage[shortlabels]{enumitem}
\usepackage[draft,color=lightgray]{todonotes} 

\usepackage{url}
\definecolor{NJIT-red}{RGB}{204, 0, 0} 

\newcommand{\Par}[1]{\left( #1 \right)} 
\newcommand{\Sqbr}[1]{\left[ #1 \right]} 
\newcommand{\Ts}{T_{\rm s}}

\begin{document}

\title[Derivative Thermodynamic Properties of Confined Fluids]{Density Functional Theory Predictions of Derivative Thermodynamic Properties of a Confined Fluid}
\author{Gennady Y. Gor}
\affiliation{Otto H. York Department of Chemical and Materials Engineering,\\
New Jersey Institute of Technology,\\
323 Dr. Martin Luther King Jr. Blvd, Newark, NJ 07102, USA}
\homepage{http://porousmaterials.net/}
\email{gor@njit.edu.}
\author{Geordy Jomon}%
\affiliation{Otto H. York Department of Chemical and Materials Engineering,\\
New Jersey Institute of Technology,\\
323 Dr. Martin Luther King Jr. Blvd, Newark, NJ 07102, USA}
\author{Andrei L. Kolesnikov}
\affiliation{Otto H. York Department of Chemical and Materials Engineering,\\
New Jersey Institute of Technology,\\
323 Dr. Martin Luther King Jr. Blvd, Newark, NJ 07102, USA}
\affiliation{Institut f\"ur Nichtklassische Chemie e. V., Permoserstra\ss{}e 15, 04318 Leipzig, Germany}

\date{\today}

\newpage

\begin{abstract}

Fluids in nanopores are of importance for many engineering applications, including energy storage in supercapacitors, hydrocarbons recovery from unconventional sources, or water desalination. Thermodynamic properties of fluids confined in nanopores differ from the properties of the same fluids in bulk.  Density functional theory (DFT) has been widely used for modeling thermodynamics of confined fluids. However, it is rarely used for calculations of derivative thermodynamic properties. Here we use a rather simple DFT model for argon based on the Percus-Yevick equation, and showed that with standard parametrization it fails to predict derivative properties. However, slight adjustment in parameters leads to quantitative predictions of isothermal compressibility and thermal expansion coefficient at a selected temperature. Using the adjusted parameterization we performed the calculations of compressibility of argon confined in carbon slit pores of various sizes, and demonstrated that the compressibility of argon in confinement is lower than that in bulk and is pore size dependent. We confirmed the DFT predictions using the Monte Carlo molecular simulations. In addition to isothermal compressibility, we calculated the thermal expansion coefficient of confined argon. Our calculations showed that it behaves similar to compressibility -- it is always lower than the bulk value and gradually increases for smaller pore sizes. For several selected pore sizes we verified the DFT calculations by Monte Carlo simulations. Overall, our results suggest that the classical DFT can be utilized for calculations of derivative thermodynamic properties of confined fluids, which are computationally challenging to  predict using molecular simulations. 
\end{abstract}

\maketitle

\section{Introduction}
\label{sec:Intro}

Derivative thermodynamic properties, such as isothermal compressibility, thermal expansion coefficient, and heat capacity, quantify the system responses to perturbations, so that they are also referred to as thermodynamic response functions. Changes in derivative thermodynamic properties are indicators of the second order phase transitions, glass transition in particular. Thermodynamic properties of fluids confined in nanopores differ from the properties of the same fluids in bulk~\cite{an2025perspective}. Most prominent examples of such differences include shifts of the phase transitions -- capillary condensation~\cite{horikawa2011capillary}, or freezing in the pores~\cite{alba2006effects}. 

Some of the derivative thermodynamic properties of fluids are also affected by confinement. Scherer, Garofalini and co-workers showed both experimentally and computationally that thermal expansion coefficient of water in nanoporous glass deviates from the bulk value, and depends on the pore size~\cite{garofalini2008molecular, xu2009thermal}. Recent experiments using ultrasound and nanoporous materials~\cite{schappert2014influence, schappert2018liquid, schappert2022distinct, schappert2024evaluating, ogbebor2023ultrasonic, karunarathne2025ultrasound, schappert2026experimental} suggested that compressibility of fluids in pores is lower than in bulk. This result was confirmed by Monte Carlo and molecular dynamics simulations~\cite{gor2015relation, dobrzanski2018effect, maximov2018molecular, corrente2020compressibility}, and recently, simulations showed that the deviation from bulk values is noticeable even for pores of tens of nanometers in size, and it gradually reaches the bulk value only at pore sizes of $\sim \SI{100}{nm}$~\cite{ogbebor2025compressibility}.

Density functional theory (DFT) has been widely used for modeling thermodynamics of confined fluids~\cite{ravikovitch2001density, de2024classical}. It is the state-of-the-art tool for prediction of adsorption isotherm for simple fluids, such as argon or nitrogen, which are used in porosimetry~\cite{landers2013density, thommes2015physisorption}. It has been routinely used for predicting adsorption in carbons~\cite{lastoskie1993pore} or silica~\cite{ravikovitch2006density} and recently has been extended to be used for MOFs~\cite{fu2015classical, stierle2024classical}. It has been also utilized for predictions of solvation pressures in the pores and adsorption-induced deformation~\cite{ravikovitch2006deformation, gor2011adsorption, gregoire2018estimation, kolesnikov2020density, dos2023adsorption, kolesnikov2024temperature}.

Several authors attempted to utilize DFT for calculating compressibility~\cite{gor2014adsorption, evans2015local, keshavarzi2016investigation, sun2019density}. Gor used quenched solid DFT~\cite{ravikovitch2006density} for argon in model silica pores, and showed the compressibility qualitatively matches experiments by Schappert and Pelster~\cite{schappert2014influence}. These results, however, showed a mismatch with the calculations from the grand canonical Monte Carlo, and in the limit of large pores did not approach the bulk value~\cite{gor2017bulk}. That mismatch posed a question, can DFT be utilized for calculations of derivative properties of confined fluids? The main goal of the current work is to address this question. We chose a simple version of DFT based on Percus-Yevick equation of state (PY EOS)~\cite{percus1958analysis}, and demonstrated that with the proper parameterization it is capable of predicting consistent derivative properties for argon in carbon pores. In addition to isothermal compressibility, we calculated thermal expansion coefficient, which also showed strong pore-size dependence.

\section{Methods}
\label{sec:Methods}

\subsection{Classical DFT}

The classical density functional theory method is widely used for modeling inhomogeneous (e.g. confined) fluids. While it has been documented in tutorial and review papers~\cite{santos2024comparison, landers2013density}, it has multiple variations in implementation, so here we present the main equations used in this work. The essence of the method is in representing thermodynamic properties as functionals of the density profile of the fluid $\rho(\mathbf{r})$, where $\mathbf{r}$ represents the spatial variable. The grand thermodynamic potential is given as
\begin{eqnarray}
    \Omega[\rho]=\mathscr{F}[\rho]+\int {\rm d} \mathbf{r} \rho(\mathbf{r}) V_{\rm ext}(\mathbf{r})-\mu\int {\rm d} \mathbf{r} \rho(\mathbf{r})
\end{eqnarray}
where $\mathscr{F}$ is the fluid free energy, $V_{\rm ext}(\mathbf{r})$ is the external potential, representing a pore, and $\mu$ is the chemical potential of the fluid. The equilibrium profile of the fluid at each chemical potential is calculated as:
\begin{eqnarray}
    \frac{\delta\Omega[\rho]}{\delta\rho(\mathbf{r})}\bigg|_{\rho_{{0}}}=0.
\end{eqnarray}

\subsection{Bulk Fluid}

Various implementations of classical DFT differ primarily in how the bulk fluid is represented. Here we follow a classical version based on the Percus–Yevick (PY) EOS~\cite{percus1958analysis}, used e.g. in \cite{ravikovitch2001density}. For calculations we utilize the numerical implementation from Ref.~\cite{kolesnikov2024temperature}. The free energy density of a bulk fluid is defined as
\begin{equation}
\label{f-bulk}
     f_{\rm bulk}(\rho) = \frac{F_{\rm bulk}(\rho)}{V} = k_{\rm B}T\Sqbr{\rho\ln\Par{\rho\Lambda^{3}} - \rho}+ f_{\rm HS}(\rho) 
     + \frac{\rho^{2}}{2}\int\limits_{V}\Phi_{\rm attr}(r){\rm d} r,
\end{equation}
where $f_{\rm HS}(\rho)$ is the hard sphere contribution from PY~\cite{percus1958analysis}, and $\int_{V}\Phi_{\rm attr}(r){\rm d} r$ is the integrated strength of the attractive Lennard-Jones fluid split according to Weeks-Chandler-Andersen (WCA) scheme~\cite{weeks1971role}.
\begin{equation}
    \int_{V}\Phi_{\rm attr}(r){\rm d} r = \frac{-32\sqrt{2}}{9}\pi\epsilon\sigma^{3} + \frac{16}{3}\pi\epsilon\sigma^{3}
    \Sqbr{\Par{\frac{\sigma}{r_{\rm c}}}^{3} - \frac{1}{3}\Par{\frac{\sigma}{r_{\rm c}}}^{9}}
\end{equation}
where $\sigma$ and $\epsilon$ are the LJ parameters and $r_{\rm c} = 5\sigma$ is the cutoff radius.

The chemical potential is defined as,
\begin{equation}
    \mu = \Par{\frac{\partial F}{\partial N_{i}}}_{T,V,N_{i\neq j}} = \Par{\frac{\partial f}{\partial\rho}}_{T,V},
\end{equation}
so Eq.~\ref{f-bulk} gives
\begin{equation}
    \label{mu-PY}
    \mu(\rho) = k_{\rm B}T\ln\Par{\rho\Lambda^{3}} + k_{\rm B}T\Sqbr{-\ln\Par{1-\eta}+\eta\frac{14-13\eta+5\eta^{2}}{2\Par{1-\eta}^{3}}} 
    + \rho\int\limits_{V}\Phi_{\rm attr}(r){\rm d} r.
\end{equation}
Similarly, pressure in the bulk phase is given as
\begin{equation}
    P = -\Par{\frac{\partial F}{\partial V}}_{T,N} = \rho\mu - f
\end{equation}
\begin{equation}
    \label{P-bulk}
    P(\rho) = \rho k_{\rm B} T\frac{1+\eta+\eta^{2}}{\Par{1-\eta}^{3}} + \frac{\rho^{2}}{2}\int_{V}\Phi_{\rm attr}(r){\rm d} r
\end{equation}
Here $\eta=\frac{\pi}{6}\rho d_{\rm HS}^{3}$ is the packing fraction, $d_{\rm HS}$ is the hard-sphere diameter of the reference system. We use the Barker-Henderson temperature-dependent hard-sphere diameter~\cite{barker1967perturbation},
\begin{equation}
    \label{d-HS}
    d_{\rm HS}=\sigma\frac{\eta_{1}k_{\rm B}T / \epsilon+\eta_{2}}{\eta_{3}k_{\rm B}T / \epsilon+\eta_{4}}
\end{equation}
where $\eta_{1}=0.3837$, $\eta_{2}=1.035$, $\eta_{3}=0.4249$, and $\eta_{4}=1$.

\subsection{Compressibility or Bulk Modulus}

Isothermal compressibility is defined as 
\begin{equation}
    \beta_{T}= -\frac{1}{V} \Par{\frac{\partial V}{\partial P}}_{T}. 
    \label{beta}
\end{equation}
However, it is often more convenient to discuss the reciprocal property --  the isothermal bulk modulus, defined as,
\begin{equation}
    K_{T}= -V\Par{\frac{\partial P}{\partial V}}_{T}.
    \label{K}
\end{equation}
When the pressure $P$ in the pore is understood as a scalar thermodynamic variable, we can relate it to the chemical potential $\mu$ of the fluid by the Gibbs-Duhem equation at constant $T$~\cite{gor2014adsorption}:
\begin{equation}
    K_{T}= \rho^{2}\Par{\frac{\partial\mu}{\partial\rho}}_{T}
    \label{K-T-density}
\end{equation}
For bulk fluid described by Eq.~\ref{f-bulk}, Eq.~\ref{K-T-density} gives
\begin{equation}
    K_{T}(\rho)= \rho k_{\rm B}T + \rho k_{\rm B}T\eta\Sqbr{\frac{1}{1-\eta}+\frac{7+\eta+\eta^{2}}{\Par{1-\eta}^{4}}} + \rho^{2}\int\limits_{V}\Phi_{\rm attr}(r){\rm d} r.
\end{equation}

\subsection{Thermal expansion coefficient}

The thermal expansion coefficient is defined as 
\begin{equation}
    \label{alpha-def}
    \alpha = \frac{1}{V}\Par{\frac{\partial V}{\partial T}}_{P},
\end{equation}
and measures the fractional change in size per degree change in temperature at a constant pressure, such that lower coefficients describe lower propensity for change in size. Eq.~\ref{alpha-def} can be rewritten as 
\begin{equation}
    \label{eq:alpha_number_density}
    \alpha = -\frac{1}{\rho}\Par{\frac{\partial \rho}{\partial T}}_{P}.
\end{equation}
It is also related to bulk modulus, such that,
\begin{equation}
    \label{eq:alpha_bulk_modulus}
    \alpha = \frac{1}{K_{T}}\Par{\frac{\partial P}{\partial T}}_{V}.
\end{equation}
Taking the temperature derivative in Eqs.~\ref{P-bulk} and \ref{d-HS}, Eq.~\ref{eq:alpha_bulk_modulus} gives for a bulk fluid:
\begin{equation}
    \alpha (\rho) = \frac{1}{K_T} \biggl(\rho\frac{1 + \eta + \eta^2}{(1-\eta)^3} + 3\rho\eta\frac{4+4\eta+\eta^2}{(1-\eta)^4}\frac{-0.0561~T/\epsilon}{(0.4249~T/\epsilon + 1)/(0.3837~T/\epsilon + 1.035)}\biggl)
\end{equation}

\subsection{Slit Pore Model}

Here we consider slit pores, being the simplest pore model, yet representing graphitic structure of nanoporous carbons. Solid-fluid interactions with each of the graphitic slabs are described within the Steele 10-4-3 potential~\cite{steele1973physical}:
\begin{equation}
\label{Steele}
\phi_{\rm sf}(z) = 2 \pi \epsilon_{\rm sf} \rho_s \sigma_{\rm sf}^2 \Delta \left[ \frac{2}{5} \left( \frac{\sigma_{\rm sf}}{z}\right)^{10} - \left( \frac{\sigma_{\rm sf}}{z}\right)^{4} - \frac{\sigma_{\rm sf}^4}{3 \Delta (z + 0.61 \Delta)^3} \right]
\end{equation}
$\Delta$ is the spacing between the graphitic slabs. And therefore the external potential is given by 
\begin{equation}
\label{Vext}
V_{\rm ext}(z) = \phi_{\rm sf}(z) + \phi_{\rm sf}(H-z)
\end{equation}

The mean pore fluid density, $n \equiv N/V$, is given by
\begin{equation}
\label{n}
n = \frac{1}{H} \int\limits_0^H \rho(z) {\rm d}z.
\end{equation}

For inhomogeneous fluid, the bulk modulus can be calculated similarly to Eq.~\ref{K-T-density}, 
\begin{equation}
    K_{T}=n^{2}\Par{\frac{\partial\mu}{\partial n}}_{T}.
    \label{K-n}
\end{equation}
The dependence $n = n(\mu)$ gives the adsorption isotherm, so Eq.~\ref{K-n} relates the modulus directly to the slope of the adsorption isotherm. However, Eq.~\ref{K-T-density} can be considered as approximate for fluids in nanopores~\cite{gor2014adsorption}.

The thermal expansion coefficient for the fluid in the pore is calculated using the same expression as calculations from the experimental data\cite{xu2009thermal}:
\begin{equation}
    \label{alpha-expt}
    \alpha = \frac{1}{V_{\rm f}}\frac{\Delta V_{\rm f}}{\Delta T},
\end{equation}
where $V_{\rm f}$ is the molar volume. We calculated $\alpha$ by using a two point forward derivative by changing temperature by $\Delta T = \SI{0.01}{K}$ at constant bulk pressure.

\subsection{Molecular Simulation}

Using statistical mechanics, the isothermal compressibility can be calculated in the grand canonical Monte Carlo (GCMC) ensemble using fluctuations of the number of particles $N$~\cite{coasne2009effect, gor2015relation, dobrzanski2021elastic}, 
\begin{equation}
    \label{comp_GCMC}
    \beta_{T}^{-1} = K_{T} = \frac{k_{\rm B} T \langle N\rangle^2}{V \langle \delta N^2 \rangle},
\end{equation}
where the temperature $T$ and volume $V$ are fixed within the ensemble. Similarly, the thermal expansion coefficient can be calculated using fluctuations of $N$ and the total potential energy of the system $U$~\cite{allen_computer_2017}.
\begin{equation}
    \label{alpha_GCMC}
    \alpha = \frac{\langle P \rangle\beta_{T}}{T}-\frac{\langle \delta U \delta N \rangle}{\langle N \rangle k_{\rm B}T^{2}} +\frac{\langle U\rangle\langle \delta N^{2}\rangle}{\langle N\rangle^{2}k_{\rm B}T^{2}},
\end{equation}
where $\langle P \rangle$ is the mean of the instantaneous pressure and $\beta_{T}$ is obtained from Eq.~\ref{comp_GCMC}.

The slit pore was modeled using the Steele potential~\cite{ogbebor2025compressibility, siderius_extension_2011} along the $z$ axis, while the $x$ and $y$ axes had a length twice the cutoff radius. Periodic boundary conditions were applied along the $x$ and $y$ axes. The initial number of particles in the system was set according to the bulk liquid density obtained from Peng-Robinson EOS. The chemical potential fixed was obtained from the same. Fluid-fluid interactions were modeled using a truncated 12-6 Lennard-Jones potential without the use of tail corrections. The system was equilibrated for $2\times10^{4}$ sets and production runs were performed for $10^{6}$ sets. Each set consisted of $10^{3}$ MC steps. There was an equal probability of MC displacement and insertion/deletion attempts. An in-house code was used to perform these simulations~\cite{ogbebor2025compressibility}. The interaction parameters for GCMC simulations are given in Table~\ref{tab:parameters}.

\section{Results}
\label{sec:Results}

\subsection{Model Parameterization}

The fluid-fluid interaction parameters for DFT models are typically found from the fit of the bulk coexistence densities, saturation pressure, and surface tension at the normal boiling point~\cite{ravikovitch2006density}. 
At VLE, the gas and the liquid phase exist simultaneously such that
\begin{eqnarray}
    \mu\Par{\rho_{\rm vap}} = \mu\Par{\rho_{\rm liq}} \\
    P \Par{\rho_{\rm vap}} = P \Par{\rho_{\rm liq}}
\end{eqnarray}

As a starting point we used the parameters for the fluid (argon)  from the DFT model by Ravikovitch and Neimark~\cite{ravikovitch2006density}, $\sigma = \SI{0.3358}{nm}$ and $\epsilon = \SI{111.95}{K}$. These parameters provide excellent agreement for vapor and liquid densities at the normal boiling point. However, they result in a significant mismatch for compressibility of bulk argon. Therefore, we took a different approach here. We optimize the fluid-fluid interaction parameters so that they provide the best values not only for liquid densities, but also for bulk modulus and thermal expansion coefficient. Given the simplicity of the chosen model we were not able to find parameters that provide agreement in a broad range of temperatures, but chose a single temperature instead, $T_{\rm s} = \SI{128.75}{K}$. Table~\ref{tab:parameters} shows the resulting parameters compared to the parameters that are frequently used~\cite{ravikovitch2006density}. The difference between the two sets of parameters is only slight, but the adjusted parameters give less than 1\% error for liquid density, and less than 4\% error for bulk modulus and thermal expansion coefficient. Note that the solid-fluid interaction parameters for argon-carbon are also different from what was established in the adsorption literature \cite{cychosz2012characterization}, as they are obtained from the Lorentz-Berthelot rules from the adjusted fluid-fluid interaction parameters.

\begin{table}[h!]
    \centering
    \begin{tabular}{|c|c|c|c|}
    \hline
    Parameter & DFT Literature & DFT This work & Monte Carlo \\
    \hline
    $T$     & \SI{87.3}{K} & \SI{128.75}{K} & \SI{128.75}{K} \\
    \hline
     $\sigma$ & \SI{0.3358}{nm} & \SI{0.3289}{nm} & \SI{0.3405}{nm} \\
     \hline
     $\epsilon/k_{\rm B}$ & \SI{111.95}{K} & \SI{111.41}{K} & \SI{119.8}{K} \\ 
     \hline
     $r_{\rm c}$ & $5 \sigma$ & $5 \sigma$ & \SI{1.4}{nm}\\
     \hline
     $d_{\rm HS}$ & $\sigma$ & Eq.~\ref{d-HS} & ~ \\
     \hline
     $\sigma_{\rm sf}$ & \SI{0.33525}{nm} & \SI{0.33445}{nm} & \SI{0.34025}{nm} \\
     \hline
     $\epsilon_{\rm sf}/k_{\rm B}$ & \SI{57.942}{K} & \SI{55.95}{K} & \SI{57.917}{K} \\
     \hline
     $\rho_{\rm s}$ & \SI{114}{nm^{-3}} & \SI{114}{nm^{-3}} & \SI{114}{nm^{-3}} \\
     \hline 
     $\Delta$ & \SI{0.335}{nm} & \SI{0.335}{nm} & \SI{0.335}{nm} \\ 
     \hline 
    \end{tabular}
    \caption{DFT parameters used for modeling argon in carbon pores in Ref.~\cite{ravikovitch2006density} and in this work. The solid fluid parameters for our work were obtained from mixing rules from C-C interaction parameters in Ref~\cite{siderius_extension_2011}. The fluid fluid parameters and the solid solid parameters used in molecular simulations are from Ref~\cite{tretiakov2004thermal} and~\cite{siderius_extension_2011} respectively.}
    \label{tab:parameters}
\end{table}

\begin{figure}[h]
\centering
\includegraphics[width=0.45\linewidth]{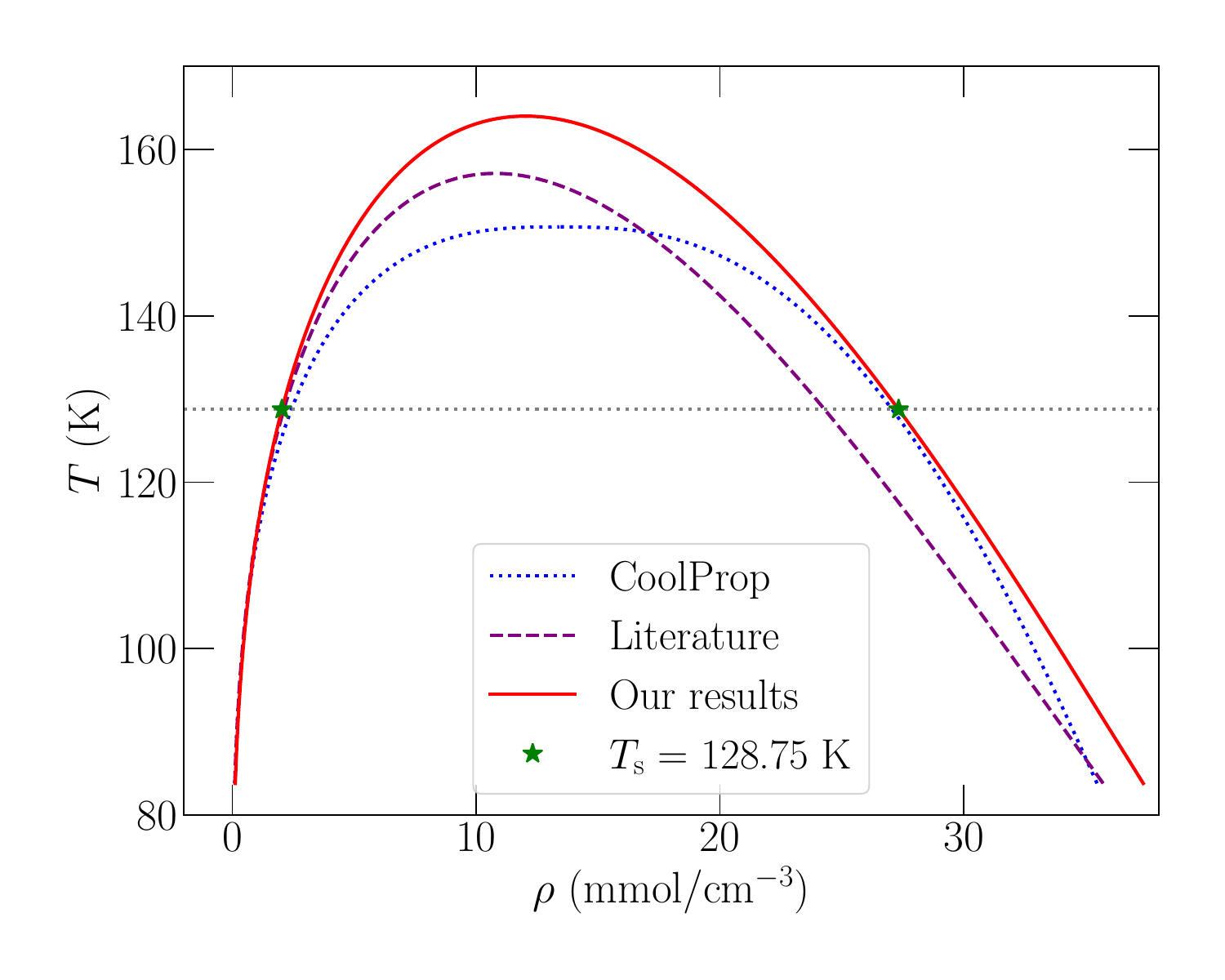}
\includegraphics[width=0.45\linewidth]{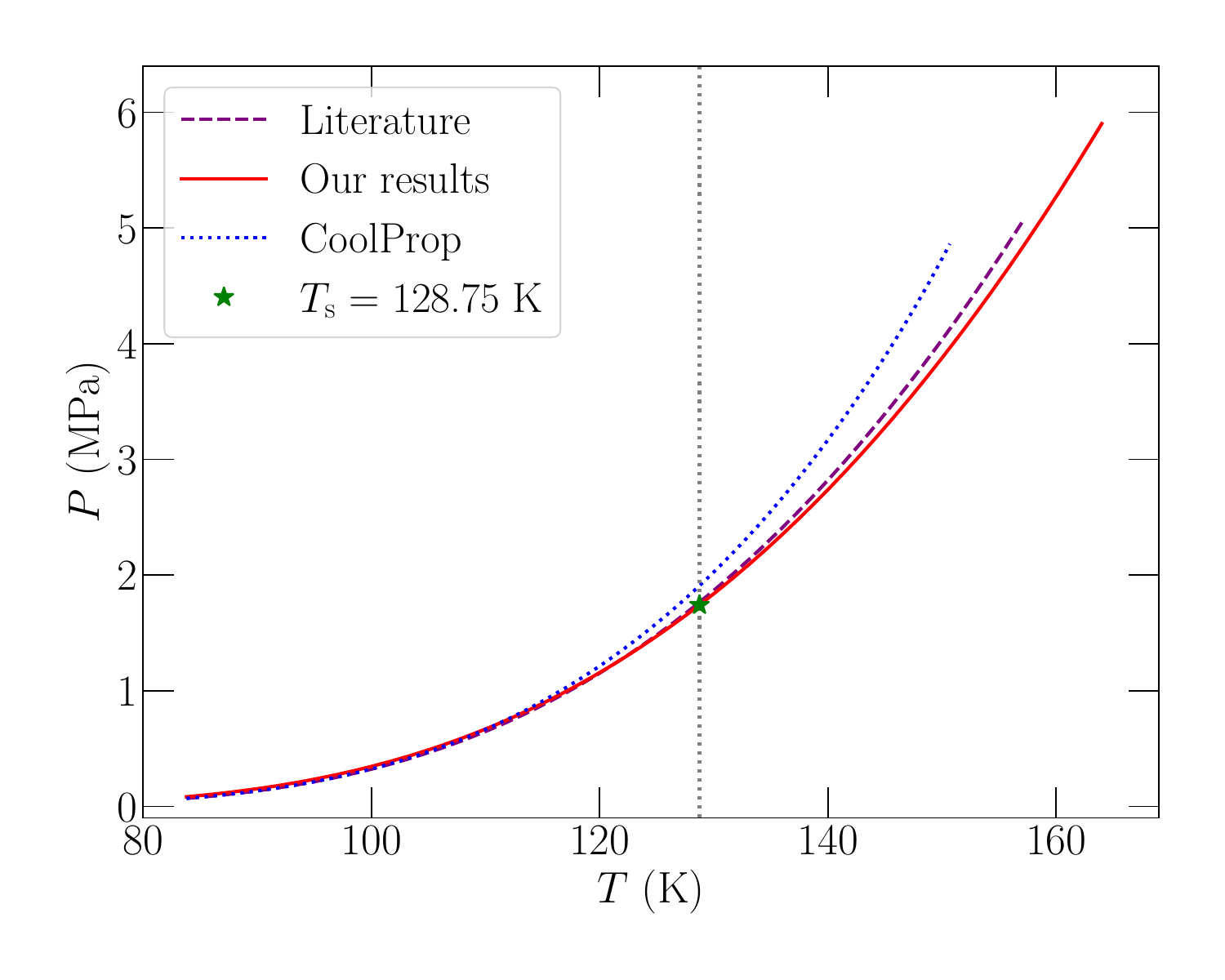}
\caption{(a) Vapor-liquid binodals for argon, dotted line is based on experimental data from CoolProp~\cite{bell2014pure}, and theoretical binodals predicted using the EOS with two different parameterizations -- from the literature~\cite{ravikovitch2006density} (dashed line) and parameterization proposed here (solid line). (b) Saturation pressure of argon as a function of temperature predicted with PY EOS and from CoolProp~\cite{bell2014pure}.}
\label{fig:binodal} 
\end{figure}

\begin{figure}[h]
\centering
\includegraphics[width=0.45\linewidth]{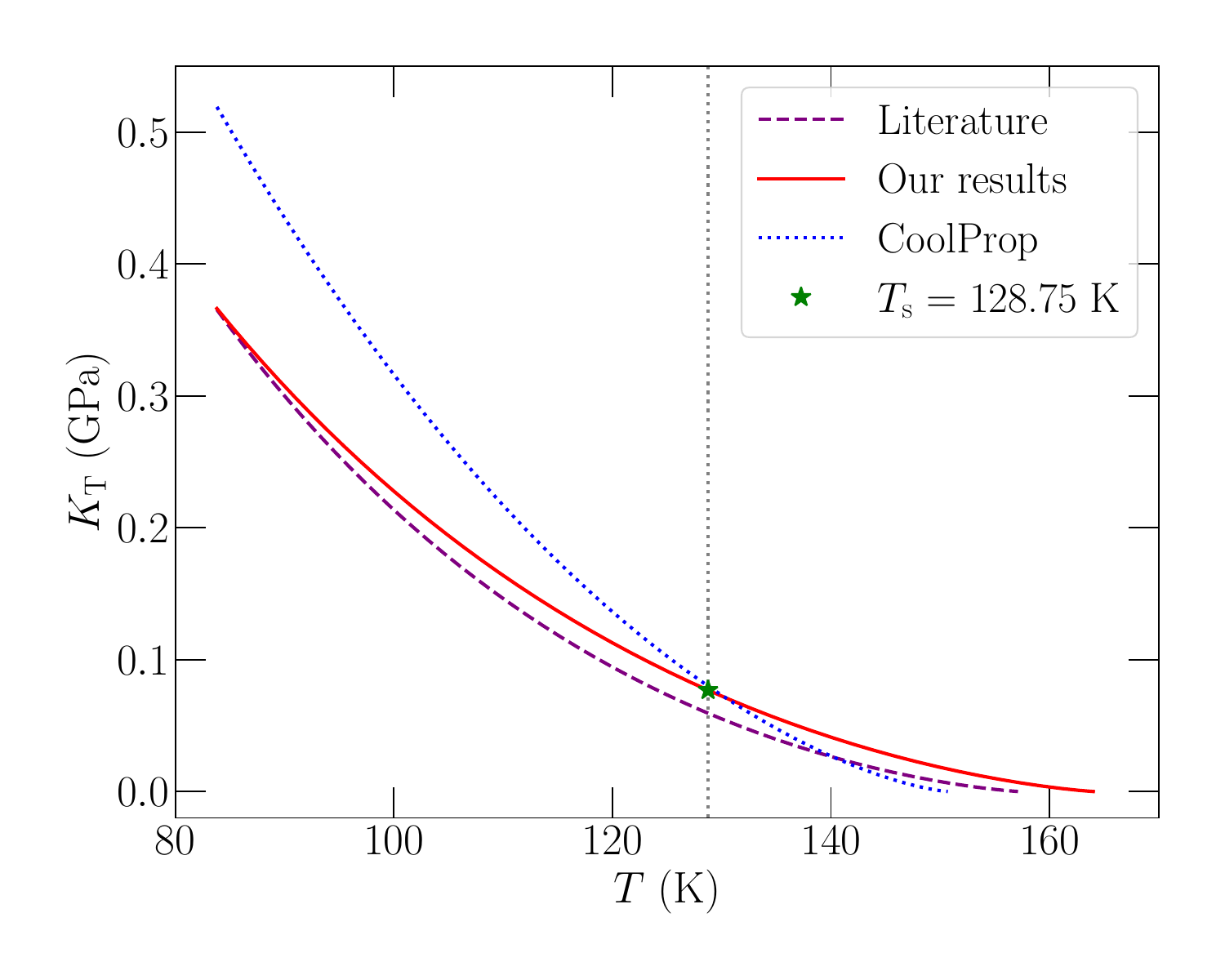} 
\includegraphics[width=0.45\linewidth]{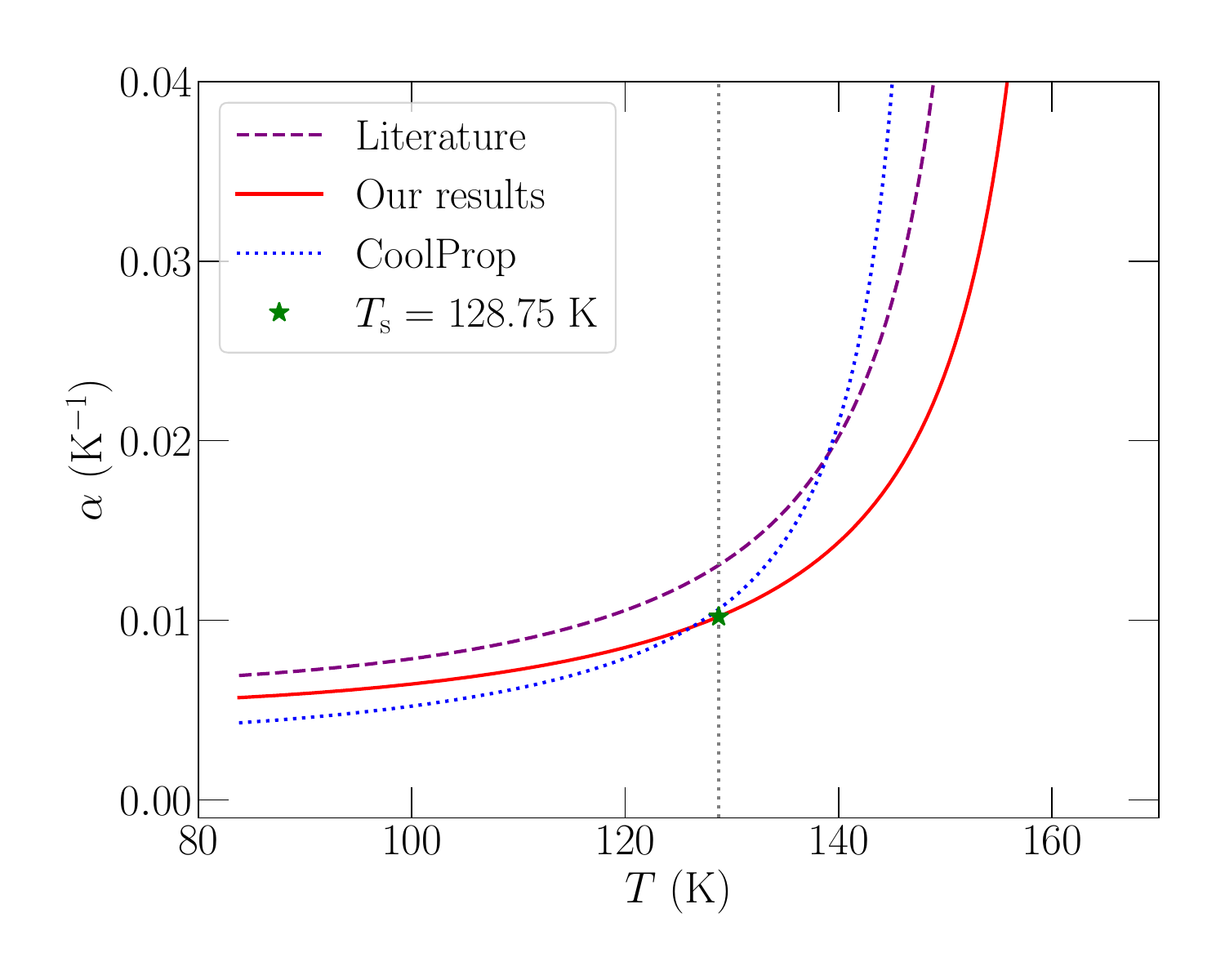} \\
\caption{(a) Bulk modulus of bulk liquid argon. (b) Thermal expansion coefficient of bulk liquid argon. Dotted line is based on experimental data from CoolProp~\cite{bell2014pure}, and theoretical binodals predicted using the EOS with two different parameterizations -- from the literature~\cite{ravikovitch2006density} (dashed line) and parameterization proposed here (solid line).}
\label{fig:bulkbulkmodulus} 
\end{figure}

Figures~\ref{fig:binodal} and \ref{fig:bulkbulkmodulus} show the thermodynamic properties of bulk liquid argon described by the PY EOS using the parameterization from the literature~\cite{ravikovitch2006density} and the adjusted parameters proposed in this work. The data is shown along with the reference data calculated using CoolProp~\cite{bell2014pure}. Figure~\ref{fig:binodal} shows the vapor-liquid binodals and the vapor pressure as a function of temperature. Figure~\ref{fig:bulkbulkmodulus} gives the bulk modulus and thermal expansion coefficients. In all of the figures, the temperature $\Ts = \SI{128.75}{K}$ is marked with asterisks. At this temperature the deviation between theoretical and experimental curves is minimal, so this temperature is chosen for all further calculations -- predictions of properties of confined argon. The literature parameterization provides a close match between theoretical and experimental values for both bulk modulus and thermal expansion coefficient at \SI{140}{K}. However, the EOS significantly underestimates saturation liquid density at the same temperature, see Fig. \ref{fig:binodal}.

\begin{figure}[h]
\centering
\includegraphics[width=0.45\linewidth]{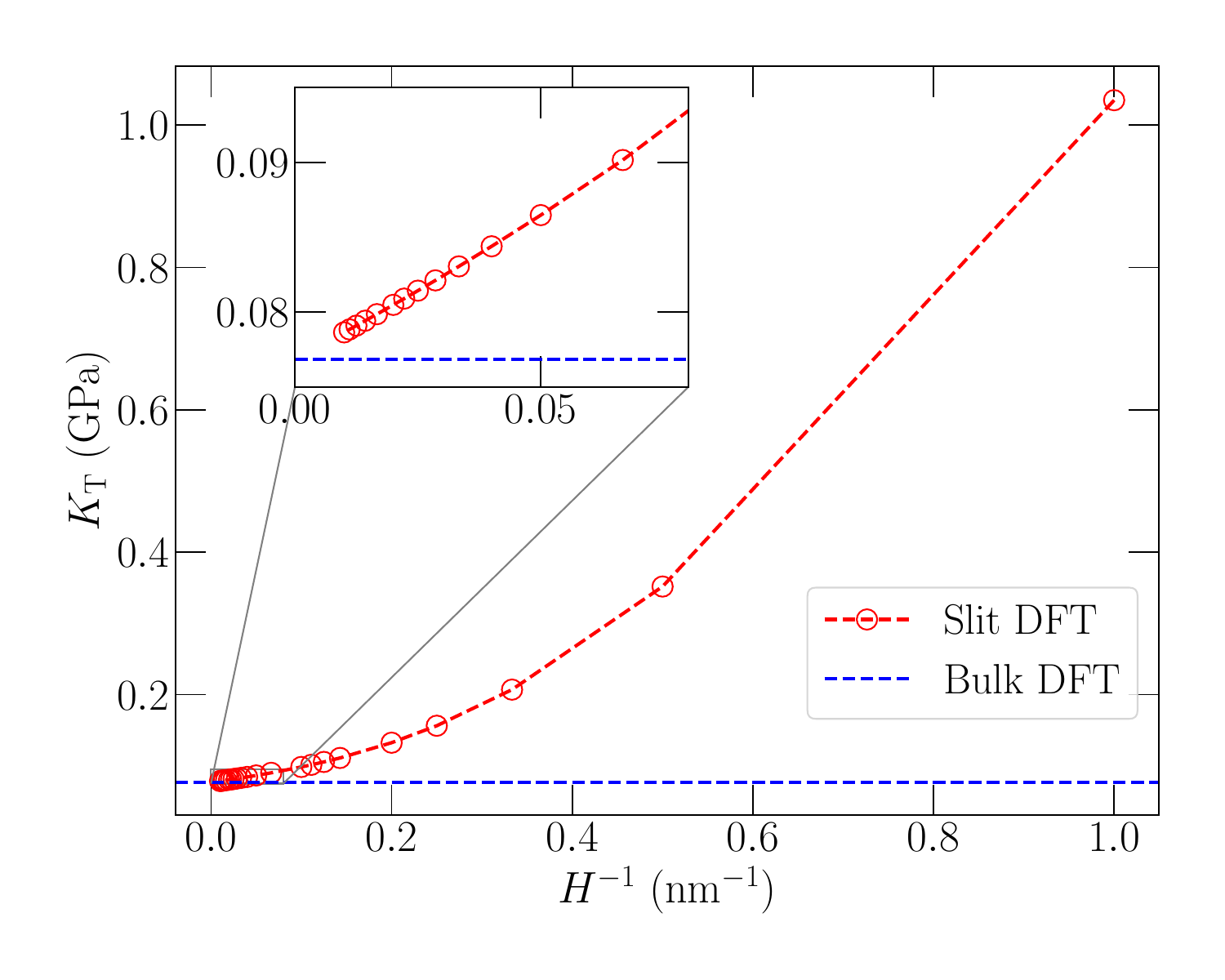}
\includegraphics[width=0.45\linewidth]{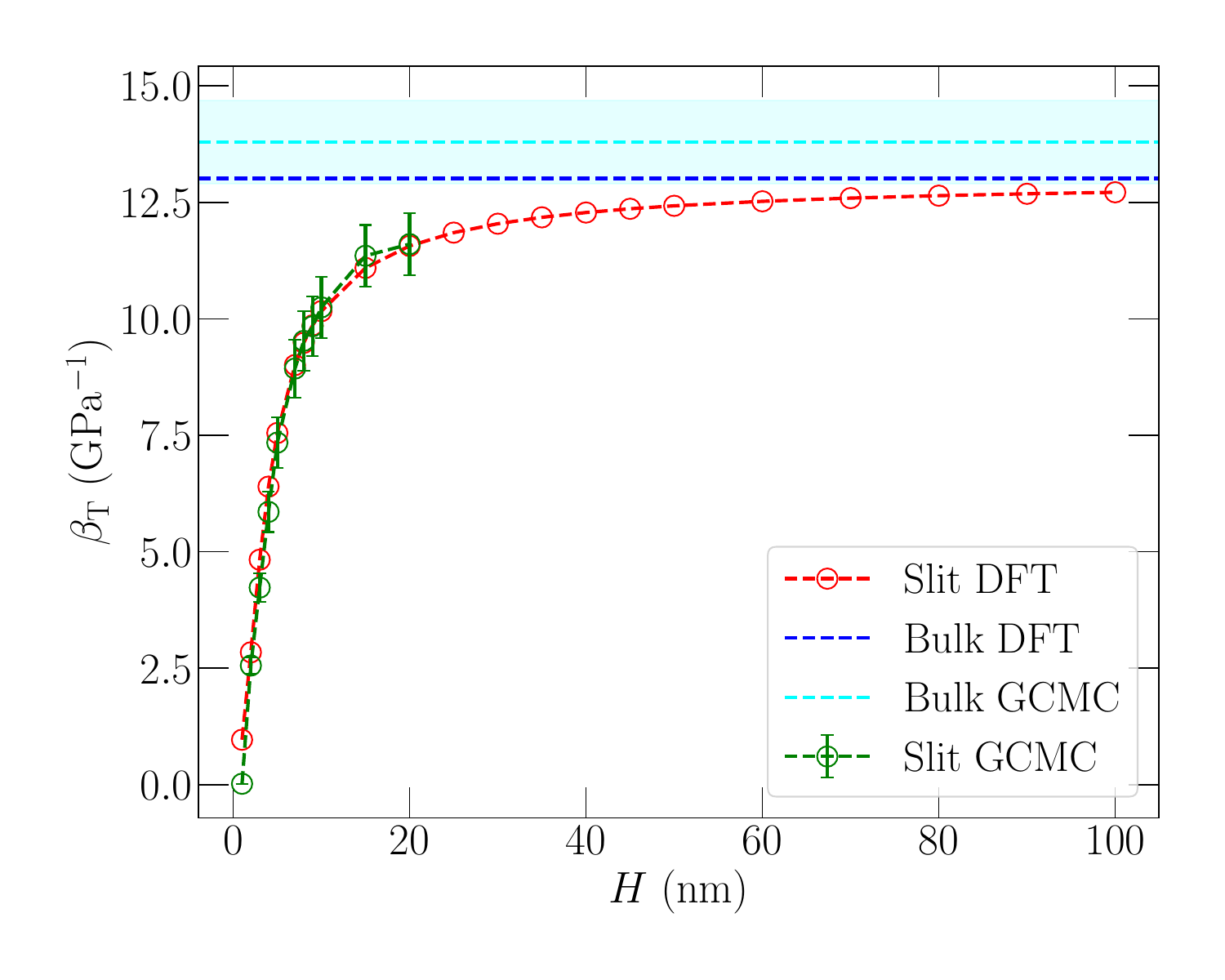}
\caption{Bulk modulus (a) and compressibility (b) of argon in slit pores of various sizes. The shaded region in Fig. ~\ref{fig:modulus-slit}b shows the error for prediction of bulk compressibility. The calculations based on DFT are in excellent agreement with the calculations from GCMC.}
\label{fig:modulus-slit} 
\end{figure}

Figure~\ref{fig:modulus-slit}a shows the bulk modulus of confined argon calculated using Eq.~\ref{K-n} for pores ranging from 1~nm to 100~nm. It shows near linear increase of the modulus with the reciprocal pore size, consistent with earlier theoretical~\cite{gor2014adsorption, corrente2020compressibility, ogbebor2025compressibility} and experimental~\cite{schappert2026experimental} studies. Figure~\ref{fig:modulus-slit}b represents the same data, plotted as $\beta_T$, which shows a gradual increase of compressibility with the increase of the pore size, nearly reaching the bulk value at $H = \SI{100}{nm}$. In addition to the data calculated based on DFT, for a limited number of pore sizes, we calculated the compressibility using Monte Carlo simulations. The results of both methods match nicely. 

\begin{figure}[h]
\centering
\includegraphics[width=0.45\linewidth]{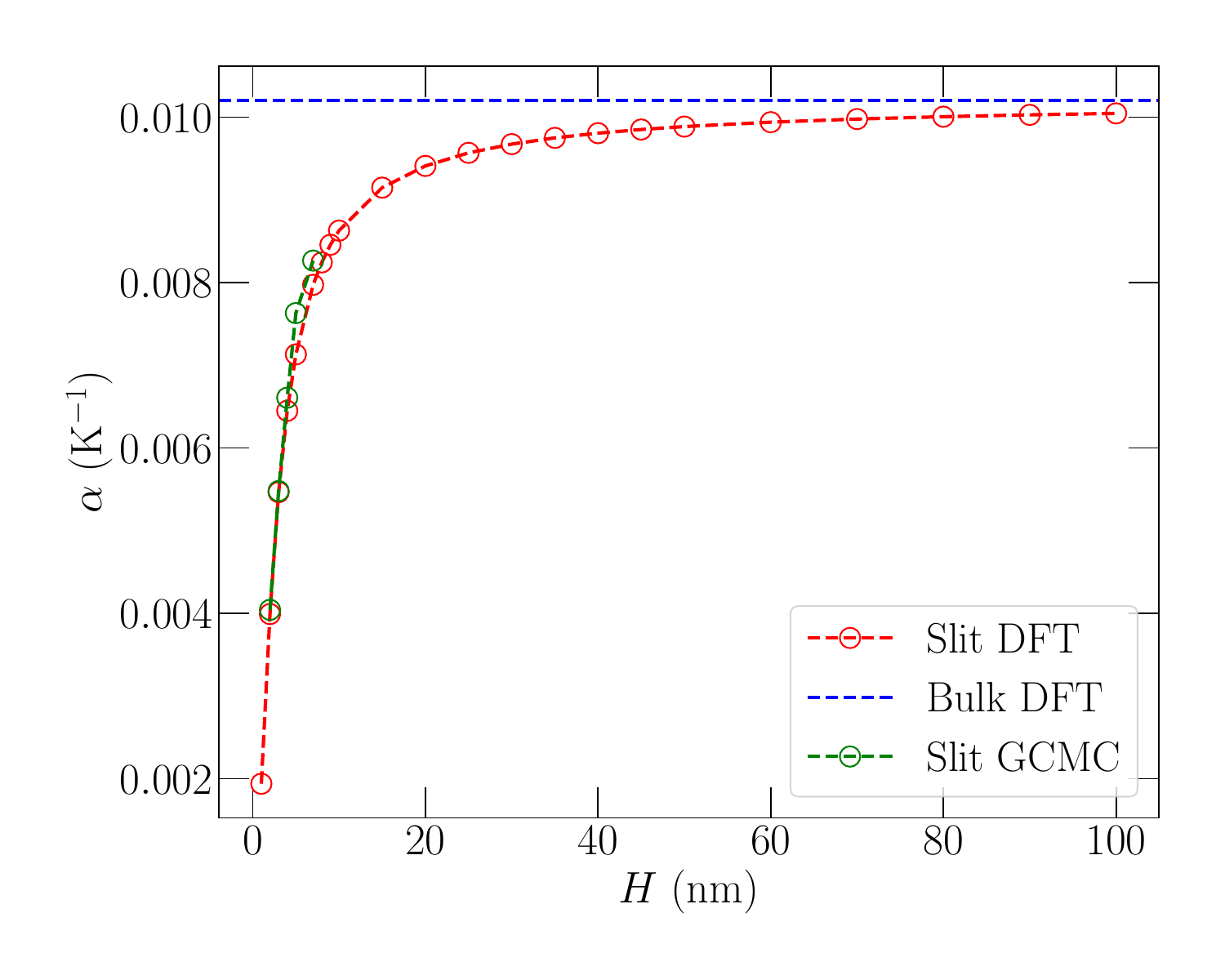}
\caption{Thermal expansion coefficient of confined argon in slits of various sizes. Calculations based on DFT (Eq.~\ref{alpha-expt}) confirmed by the calculations based on molecular simulations (Eq.~\ref{alpha_GCMC}).}
\label{fig:alpha-slit} 
\end{figure}

Figure~\ref{fig:alpha-slit} shows thermal expansion coefficient of confined argon in slits of various sizes, calculated using Eq.~\ref{alpha-expt}. It shows the significant deviation from bulk values at pore sizes below 10~nm, and gradual increase towards the bulk limit at approximately 100~nm. In addition to the DFT calculations, the figure shows the results of calculations using Monte Carlo simulations for selected pore sizes (\SI{2}{nm} to \SI{7}{nm}).  

\section{Discussion}
\label{sec:Discussion}

While our study does not aim to quantitatively describe experimental results, it is worth mentioning, that experiments suggested the deviations of thermal expansion coefficient of fluids in the pores from the bulk value~\cite{alba2003confinement, morineau2004structure}. Thermal expansion coefficient of benzene in pores of a few nm in size are noticeably lower than thermal expansion coefficient of bulk benzene, which is consistent with our predictions. 

Interestingly, the deviation of thermal expansion coefficient for water in silica pores, showed a different trend compared to argon -- thermal expansion coefficient of confined water appeared higher than for bulk, and for smaller pores the deviation was higher~\cite{xu2009thermal}. However, water is known to have many anomalies, including the anomaly of thermal expansion coefficient, and this effect is worth a separate study.

Calculations of derivative properties of fluids using molecular simulations is computationally expensive~\cite{ogbebor2025compressibility}. Our results showed that with a proper EOS and parameterization, DFT can offer a fast alternative to molecular simulations, accurately predicting the isothermal compressibility and thermal expansion coefficient. This work can be extended further using more advanced DFT models such as the three dimensional models, which can represent complex pore geometries~\cite{bernet20203d} \cite{santos2024comparison, dos2025slit}. Another possible extension would be using DFT, based on more advanced EOS, such as based on the PC-SAFT EOS~\cite{lu2023dualistic, mayer2024adsorption, teh2025classical}.

\section{Conclusion}
\label{sec:Conclusion}

Experiments show that derivative thermodynamic properties, compressibility and thermal expansion coefficient, of confined fluids differ from these properties in bulk. Molecular simulations confirm these deviations. However, earlier attempts to use classical DFT fail to quantitatively reproduce compressibility of even a fluid as simple as argon. Here we utilized a simple version of DFT, based on the Percus-Yevick EOS, and showed that slight adjustment of the fluid-fluid interaction parameters helps to reproduce the compressibility and thermal expansion coefficient of bulk liquid argon at a selected temperature. We then used this parameterization and showed that it allows to obtain the compressibility and thermal expansion coefficient of argon in carbon slit pores, which match quantitatively with the calculations based on established molecular simulation techniques. We show that both properties are pore-size dependent, approaching the bulk values for pore sizes of 100 nm. Given the low computational costs of classical DFT, our work provides an opportunity for calculation of derivative thermodynamic properties of fluids in larger pores within reasonable computational effort.

\section*{Acknowledgments}

This work was supported by the National Science Foundation (grant CBET-2344923). G.J. thanks the support of the NJIT Provost Undergraduate Research and Innovation (URI) Summer Fellowship, sponsored by PSEG. The authors thank Santiago Flores Roman for help with Monte Carlo simulations.

\section*{Data Availability Statement}

The data that support the findings of this study are available within the article. 

\clearpage

\bibliography{dft}
\bibliographystyle{apsrev}

\end{document}